%% file: sprintaarevised.tex
\documentclass[12pt]{siamltex}
\usepackage{amsmath,amssymb,epsfig,graphicx}

\usepackage{hyperref,subfigure}
\usepackage{color}

\if {

\oddsidemargin=0cm
\topmargin=-1cm
\textheight=24.2cm
\textwidth=15cm

\theoremstyle{plain}
\newtheorem{theorem}{Theorem}[section]

\theoremstyle{remark}
{

}
} \fi

\newtheorem{remark}[theorem]{Remark}

\input{macro}
\def\beq{\begin{equation}}
\def\eeq{\end{equation}}
\def\be{\begin{equation}}
\def\ee{\end{equation}}

\begin{document}

\title{How to run  100 meters?}

\author{Amandine Aftalion\thanks{Ecole des Hautes Etudes en Sciences Sociales, PSL Research University, 
CNRS UMR 8557, Centre d’Analyse et de Math´ematique Sociales, 54 Boulevard Raspail, Paris, France;
amandine.aftalion@ehess.fr}}
\date\today
\maketitle
\date\today

\begin{abstract} The aim of this paper is to bring a mathematical justification to the optimal way
 of organizing one's effort when running. It is well known from physiologists that all running exercises of
  duration less than 3mn are run with a strong initial acceleration and a decelerating end; on the contrary, long
  races are run with a final sprint. This can be explained using a mathematical model describing the evolution
   of the velocity, the anaerobic energy, and the propulsive force:
    a system of ordinary differential equations,
     based on Newton's second law and energy conservation,  is coupled to the condition of
     optimizing the time to run a fixed distance. We show that the monotony of the velocity curve vs time
     is the opposite of that of the oxygen uptake ($\dot{VO2}$) vs time. Since the oxygen uptake is monotone increasing
      for a short run, we prove that the velocity is exponentially increasing to its maximum and then decreasing.
      For longer races, the oxygen uptake has an increasing start and a decreasing end and this accounts
      for the change of velocity profiles. Numerical simulations are compared to timesplits from real
    races in world championships for $100$m, $400$m and $800$m and the curves match quite well.

\end{abstract}


\pagestyle{myheadings}
\thispagestyle{plain}
\markboth{A. Aftalion}{How to run a 100m?}

\renewcommand{\thefootnote}{\fnsymbol{footnote}}



\section{Introduction}\label{introsec}

When watching a 100m race in a world championship or the Olympic games, one is not always aware
 that runners do not finish the race speeding up but slowing down.
 More precisely, they accelerate for the first 70m and then slow down in the last 30m.
  This can be checked by looking at the time splits every 10m for all athletes. Table \ref{tab1}
  provides an example for the winner of the World Championship in 2011. One can notice that at 70m, the timesplit
  increases, which means that the velocity decreases.
  \begin{table}[!htb]
\caption{Blake's timesplits for the 2011 World Championships. Line 1 : distance in meter, line 2, cumulated time in seconds, line 3, time splits for 10
meters in seconds.}\label{tab1}
\begin{tabular}{ccccccccccc}
\hline\\[-1.5ex]
 10m & 20m & 30m & 40m & 50m &  60m  & {\color{red} 70m} & 80m & 90m & 100m\\[0.5ex]
\hline\\[-1.5ex]
  1.87 & 2.89 & 3.82 & 4.70 & 5.56 & 6.41 & 7.27  & 8.13 & 9.00 & 9.88 \\[0.5ex]\hline \\[-1.5ex]
 1.87 & 1.02 & 0.93 & 0.88 & 0.86 &  0.85  & {\color{red} 0.86} & 0.86 & 0.87 & 0.88\\[0.5ex]
\hline
\end{tabular}
\end{table}
   This way of running is not because they accelerate too strongly at the
    beginning of the race, or are exhausted, but because this is the best way to run
    a 100m from the physiological point of view. One of the aims of this paper is to bring a mathematical
     justification and explanation to this phenomenon, and explain why, using coupled ordinary
     differential equations,
  the best use of one's ressources leads to a run with the last third in deceleration.

  In fact all distances are not run the same way \cite{hanoneffects}: for distances up to 400m,
   the last part of the race is run slowing down, while for distances longer than 1500m,
   the first part of the race has an initial acceleration, the middle part is run at almost constant
   velocity and there is a final acceleration. The 800m is an intermediate race. This way of running is
   well known from physiologists but they do not have an explanation of why the human body
    does not optimize the same way according to the distance.
    This  paper aims at understanding this optimization according to the distance with a mathematical justification using the model developed in \cite{AB}.

Up to now, very simple mathematical models have been used in the field of sport sciences
 to model the velocity evolution in a sprint. The recently published works  \cite{sam15,rab15}
  model the velocity
  curve in a 100m as a double exponential $v(t)=v_{max} (e^{-t/\tau_1}-e^{-t/\tau_2})$,
  and they fit the parameters $\tau_1$ and $\tau_2$ to match real curves. This is indeed
   very close to the real
  velocity curves but the authors do not provide an explanation of their modelling. In this paper, we provide a model of coupled
  equations between velocity, energy and propulsive force which accounts for this double exponential approximation.

\hfill

 A pioneering mathematical work is that of Keller
\cite{Ke2}
relying on Newton's law of motion and energy conservation. Though his analysis reproduces
 quite well the record times for distances up to 10km, it does not reproduce the champions' way of running. Indeed, Keller's model relies on the assumption of constant $\dot{VO2}$,
  that is constant oxygen uptake and 
  it is proved in \cite{AB}, that the race is made up of exactly three parts:
  \begin{enumerate}\item initial speed up phase at maximal force
  \item the velocity is constant \item the velocity decreases and the runner runs with zero energy.\end{enumerate}
     Therefore, in order to find non constant velocities as in real races,
        one has to take into account a realistic $\dot{VO2}$ description.
        Several improvements of Keller's model have been introduced: the effect of fatigue \cite{Ma,Wo}, the variation in maximal oxygen uptake \cite{Be,be97}, air resistance and altitude \cite{besmall,Quinn}, track curvature \cite{besmall}. Other related works include \cite{alvarez,M3param,Mcritical,pt,WS1}.

  In \cite{AB}, a new model is introduced  relying on Keller's equations \cite{Ke2} but
  improving them using a hydraulic analogy \cite{Mcritical} and
      physiological indications \cite{hanoneffects}, and taking
       into account a realistic model for $\dot{VO2}$, the oxygen uptake.
      Numerical simulations are performed in \cite{AB} for a 1500m.
\begin{figure}
\centering
\begin{center}
\includegraphics [width=10cm]{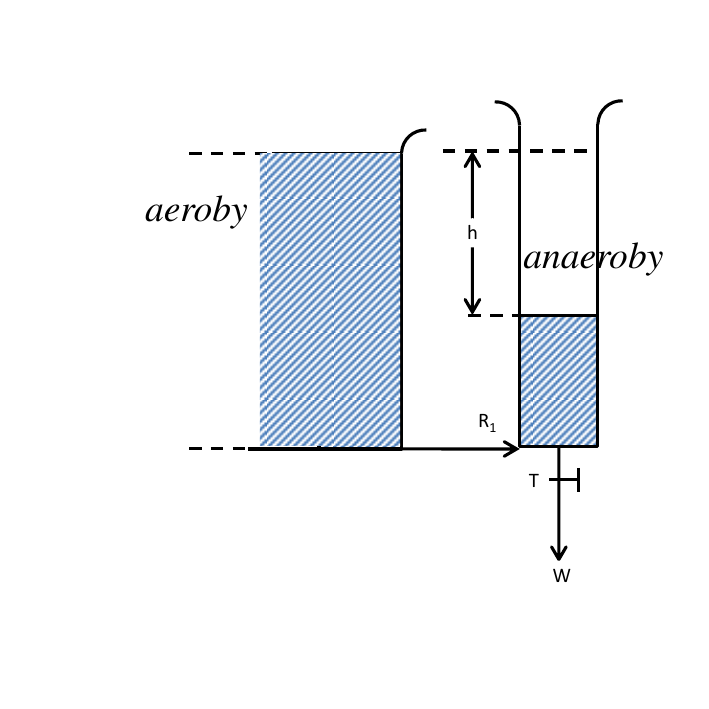}
\end{center}{\vspace{-2cm}}
\caption{Scheme of the container modeling.}
\label{figslowingh}
\end{figure}
Let us introduce the model of \cite{AB} that we adapt to short races.
 The first equation is
   the equation of motion, as in Keller's paper: \beq\label{acc} \frac{dv}{dt}+\frac v \tau =f(t)\eeq
   where $t$ is the time, $v(t)$ is the instantaneous velocity, $f(t)$ is the propulsive force and $v/\tau$ is a resistive force per unit mass. The resistive force can be modified to include another
   power of $v$ or the influence of slopes, by adding to the right hand side
 a term of the form
$-g\sin\alpha(d(t))$, where $\alpha(d(t))$ is the slope at distance $d(t)$. We can relate $\sin \alpha(d)$ to $A(d)$, the altitude of the center of mass of the runner at distance $d$, by $$\sin\alpha(d(t))=\frac{A'(d(t))}{\sqrt{1+(A'(d(t)))^2}}.$$ For most races, one can assume that
 $\sin\alpha(d(t))\sim A ' (d(t))$ and here, we assume for simplicity that $A$ is constant along the race.

   The second equation is an equation governing the energy.
    In fact, human energy can be split into aerobic energy called $e_{ae}(t)$,
    which is the energy provided by oxygen consumption, and
 anaerobic energy $e (t)$, which is provided by glycogen and lactate.
 A very good review on different types of modeling can be found in \cite{Mcritical}.
  In \cite{AB}, it is assumed that the anaerobic energy has finite capacity and is modeled by a container
   of finite height (set to 1) and section as illustrated in Figure \ref{figslowingh}.
   When it starts depleting by a height $h$, then what is called in physiology
   the accumulated oxygen deficit
   which is proportional to the height, is reduced to $e^0-e$, where $e^0$ is the initial energy.
 It is assumed that the aerobic energy is of infinite capacity and
    flows at a  rate of $ \sigma$ through a connecting tube $R_1$ into the anaerobic container. Note that $\sigma$
     is the energetic equivalent per unit time of $\dot{VO2}$,
      the volume of oxygen used by unit of time.
      This equivalent can be determined thanks to the Respiratory Exchange Ratio and
      depends on the intensity of effort.
      Nevertheless, a reasonable average value is that 1l of oxygen produces 20kJ \cite{Per}.
      The flow of the aerobic container into the anaerobic one depends on the anaerobic energy level, which is
      why we model $\sigma$ as a function of $e$: if the level of the anaerobic container is above the connecting
      tube $R_1$, as on the figure, then $\sigma$ is proportional to $h$, the
difference of fluid heights in the containers, that is $e^0-e$, while if it is
      below, then $\sigma$ is constant.

      For a short race,  $R_1$ is connected to the bottom of the anaerobic container as in Figure \ref{figslowingh}. Therefore, $\sigma$ is proportional to $e^0-e$,
       but its maximal value $\bar \sigma$ is
      not reached, so that $\sigma$ can be modelled as
      \be\label{sigmashort}
      \sigma (e)=\lambda  \bar \sigma
      (e^0-e)\ee where $\lambda$ is such that $\lambda e^0<1$.
The available flow at the bottom of the anaerobic container is the work of the
        propulsive force $f(t) v(t)$ and is equal to the creation of available energy through
         $\sigma(e)$ (aerobic energy)
        and
        $-de/dt$ (anaerobic energy). This leads to the energy equation:
     \be\label{eqsumean}\frac {d e}{dt}=\sigma (e(t)) - f(t)v(t). \ee Let us point out that $e(t)$
     is a decreasing function of time, therefore $\sigma (e(t))$ is  an increasing function of time, as expected. An example of $\sigma (e(t))$ for  a 400m is illustrated in Figure \ref{figvo2}.

     When the race is longer, the dependence of $\sigma$  on time is well known
      \cite{hanoneffects} and varies according to the length of
     the race: $\sigma$, as a function of time, is linear increasing until it
   reaches the maximal value $\bar\sigma$ and then is decreasing to its final value $\sigma_f$,
   which corresponds to the rate of aerobic transfer when there is no anaerobic energy left.
   As a function of energy, we can take the following model, which has the opposite monotony
   since $e(t)$ decreases from $e^0$ to 0 with time:
      \be\label{sigmavar}
\sigma (e)=\left\{\begin{array}{ll}
\bar \sigma \frac{e}{e^0
      e_{crit}}+\sigma_f (1-\frac{e}{e^0 e_{crit}})\hbox{ if }\frac{e}{e^0}<e_{crit}\\
\bar\sigma
    \hbox{ if }\frac{e}{e^0}\geq e_{crit} \hbox{ and }
    \lambda (e^0 -e)\geq1\\ \lambda \bar \sigma (e^0 -e)
\hbox{ if }\lambda (e^0
    -e)<1 \end{array}\right.
\ee where $e^0$ is the initial value of energy, $\bar \sigma$ is the maximal value of $\sigma$, $\sigma_f$ is the final value of $\sigma$ at the end of the race,
  $e_{crit}$ the critical energy at which the flow
 of aerobic energy into
  the anaerobic container starts to depend on the residual anaerobic energy, because of an extra retroc control mechanism. The parameters
  $\lambda$, $e^0$, $e_{crit}$, $\bar \sigma$, $\sigma_f$ depend on the runner and on the race.
   Then (\ref{eqsumean}) also holds. An example of $\sigma (e(t))$ for  a 800m are illustrated in Figure \ref{figvo2} or for a 1500m in Figure \ref{fig1500}.

Constraints have to be imposed; the force is controlled by the runner but it cannot exceed a maximal value
 $f_M$ and the energy is nonnegative:
   \beq\label{force} 0\leq f(t)\leq f_{M},\hbox{ and } e(t)\geq 0,\eeq with the initial conditions:
   \be v(0)=0,\quad e (0)=e^0.\label{initcond}\ee
     The aim is to minimize the time $T$, given the distance $D=\int_0^T v(t)\ dt$. The minimization problem
     depends on numerical parameters $\tau$, $f_M$, $e^0$, $\bar \sigma$, $\sigma_f$, $\lambda$, $e_{crit}$.

 \section{Numerical presentation of the models}

   Our
 numerical simulations are  based on the Bocop toolbox
for solving optimal control problems \cite{bocop}. This software
combines a user friendly interface, general Runge-Kutta
discretization schemes described in
\cite{MR1804658}, and the numerical resolution of the
discretized problem using the
nonlinear programming problems solver IPOPT \cite{MR2195616}.

Numerically, given the time splits of Table \ref{tab1}, Bocop identifies the parameters $\tau$, $f_M$, $e^0$, $\bar \sigma$, $\sigma_f$, $\lambda$, $e_{crit}$
 that match the time splits and
 provide the optimal velocity curve. Another protocol to identify the parameters has been described in
 \cite{aftaction}.

\begin{figure}
\centering
\begin{center}
\subfigure{\includegraphics [width=8cm]{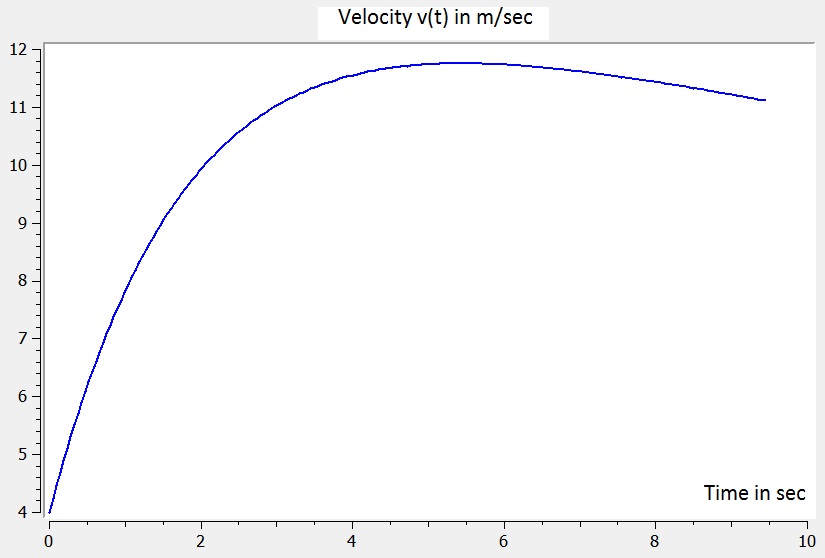}}
\quad
\subfigure{\includegraphics [width=10cm]{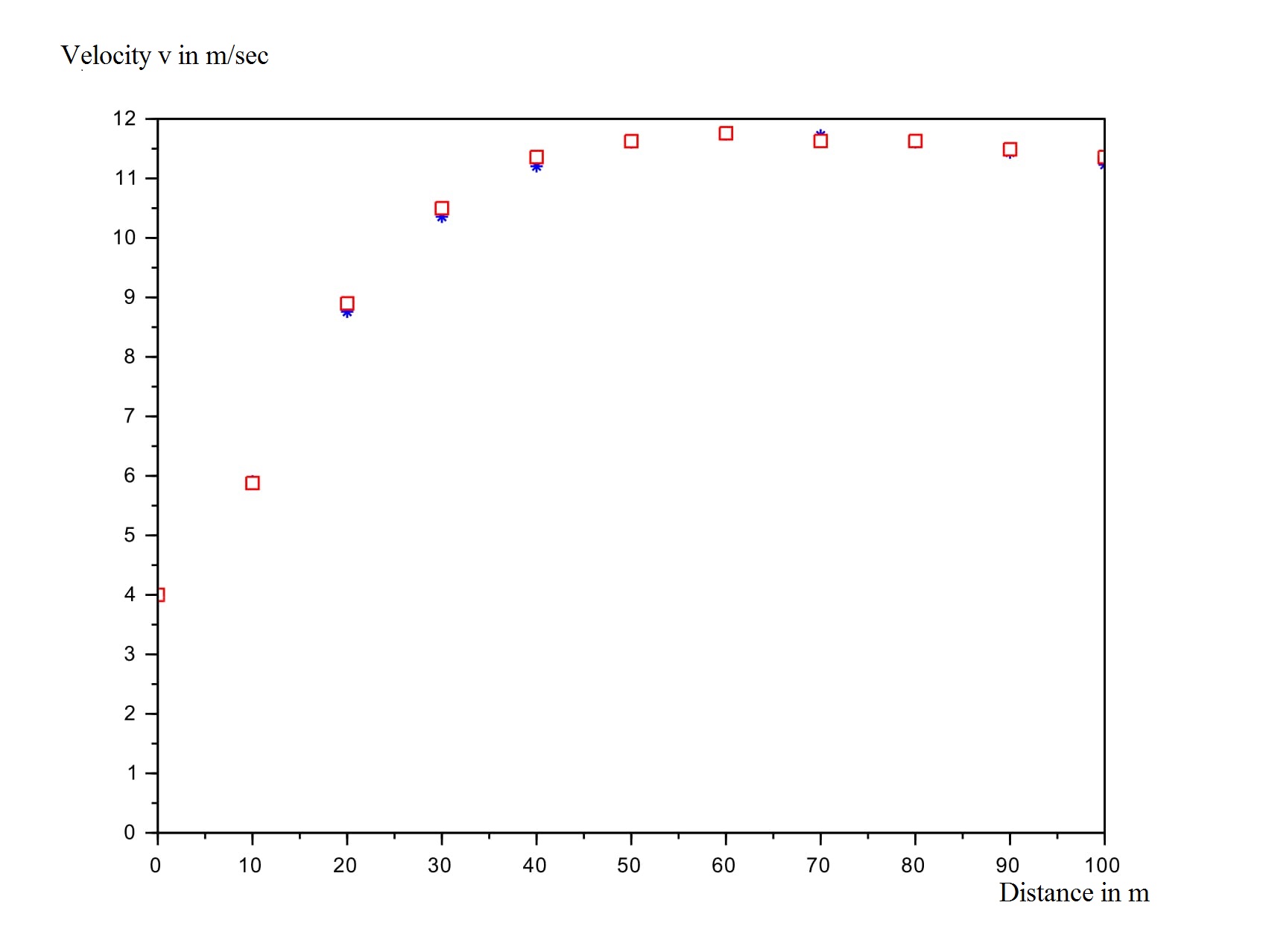}}
\end{center}
\caption{Top: Plot of the velocity $v(t)$ vs time for Blakes' 100m. Bottom: every 10m, the numerical velocity is averaged (star)
 and compared to the mean value obtained from Table \ref{tab1}, as the ratio of 10 to the timesplit (square).
 When they coincide, only the square is visible. The velocity is in
  $m.s^{-1}$ and is plotted vs the distance in m.}
\label{fig100M}
\end{figure}

 In Figure \ref{fig100M},  the numerically computed velocity is plotted as
 a function of time on the left, while, on the right, a mean value every 10m is computed (star) to compare
 with the mean value from Table \ref{tab1} (square). The matching is quite good.
 The initial velocity is taken to be 4, instead of 0, to better
 take into account the departure in the starting blocks \cite{sam15}. The left velocity curve indeed looks like a double exponential.

The rest of the paper is devoted to a mathematical justification of the race.
\section{Mathematical analysis}\label{secmathana}

We consider the following state equation
\be
\label{spmpsdrse}
\dot h(t) = v(t), \quad \dot v (t)= f(t) - v(t)/\tau,\quad \dot e(t)= \sigma(e(t)) - f(t)v(t),
\ee
We assume that the  recreation function
$\sigma(e)$ satisfies
\be
\label{spmpsre}
\text{$\sigma(e)$ is $C^2$ and nonnegative.}
\ee
We will apply it to $\sigma (e)= \alpha (e^0 -e)$ for the sprint and to a $C^2$ regularization of (\ref{sigmavar})
 for longer races.
The initial conditions are
\be
\label{introsec4}
h(0)=0, \quad v(0)=0, \quad e(0)=e^0 >0,
\ee
and the constraints are
\be
\label{introsec4cc}
0 \leq f(t)\leq f_M, \quad e(t)\geq 0, \;\; t \in (0,T),\quad
h(T) = D.
\ee
The optimal control problem is to minimize the final time:
\be
\Min T; \quad
\text{s.t. \eqref{spmpsdrse} and
\eqref{introsec4}-\eqref{introsec4cc}.}
\ee
Note that we could as well take the final constraint as
$-h(T)\leq -D$.
Writing an inequality in this way yields the sign of the
Lagrange multiplier.

It is established in \cite{AB}
that the optimal solutions (for a given distance $D$)
with corresponding time $T$
are also solutions of the problem of maximizing
the distance over a time interval $T$. So, in the
 following, we will rather use this second formulation for the proofs.

The  optimal control problem is
\be
\label{extsecseocpR1}
\Min - \int_0^T v(t) \dd t; \quad
\text{s.t. \eqref{spmpsdrse} and
\eqref{introsec4}-\eqref{introsec4cc}
hold.}
\ee
The Hamiltonian function is
\be
H[p](f,v,e) := -v + p_h v+p_v (f-v/\tau)
+ p_e (\sigma(e) - f  v).
\ee
The costate equation is therefore,
omitting time arguments:
\be
\label{costeqaR}
\left\{ \ba{lll} -\dot p_h &= 0\\
-\dot p_v &= -1 - p_v /\tau- p_e f

\\
-\dd p_e &= p_e \sigma'(e) \dd t - \dd \mu,
\ea\right.
\ee
with final conditions
\be
\label{costeqaRfc}
p_v(T) = p_e(T) =0.
\ee We can fix therefore $p_h$ to 0. The variables $p_v$ and $p_e$ are called costate variables.
 Here $\dd\mu$, identified to the bounded variation function
$\mu$ on $[0,T]$,
is a Borel measure (that can be interpreted as a
Lagrange multiplier) associated to the state constraint $-e\leq 0$ which satisifes
\be
\dd\mu \geq 0; \quad
\supp(\dd\mu) \subset \{t\in [0,T]; \; e(t)=0\}.
\ee

If $0\leq a < b \leq T$ is such that
$e(t)=0$ for $t\in [a,b]$, but $e(\cdot)$ does not vanish over
an interval in which $[a,b]$ is strictly included, then we say that
$(a,b)$ is an
{\em arc with zero energy}.
Similarly, if $f(t)=f_M$ a.e. over $(a,b)$ but not over an open
interval strictly containing $(a,b)$, we say that $(a,b)$ is an
{\em arc with maximal force}.  A {\em singular arc} is one
over which the bound constraints are not active. We recall a result from \cite{AB}:

\begin{theorem} \cite{AB}
\label{maint.t1}The above problem (\ref{extsecseocpR1}) has at least one optimal solution.
 An optimal trajectory starts with a maximal force
arc, and is such that $e(T)=0$ and $\mu$ has no jump. Moreover, $p_e(t) < 0$ and $p_v(t) < 0$
 for $t\in [0,T)$.
\end{theorem}

Numerically, the zero energy arc is only a few points and cannot be seen on a 100m.
 Note that if we plug $f(t)=f_M$ in the second equation of (\ref{spmpsdrse}), then we find as a solution
  $v(t)=f_M\tau (1-e^{-t/\tau})$.

  Once we know that the race starts with an arc of maximal force and ends with a zero energy arc,
 we want to know more about the singular intermediate arc.

\subsection{Justification of the velocity decrease in a 100m race}

In this section, we will find properties of the singular arc.  We will prove the following:
\begin{theorem}\label{theomain} Let $\sigma(e)=\alpha (e^0-e)$ with $\alpha \tau <1$. Then an optimal trajectory of (\ref{extsecseocpR1})
 starts with a maximal force arc, followed by a singular arc on which \be\label{eqvsimp}
\dot{v}+\alpha v=0\ee  and ends with a zero energy arc. Therefore, the velocity profile is
\begin{eqnarray}\label{v1} v(t)&=&f_M\tau (1-e^{-t/\tau})\hbox{ for }0\leq t\leq t_1\\
 v(t)&=&f_M\tau (1-e^{-t_1/\tau})e^{-\alpha(t-t_1)}
 \hbox{ for } t_1\leq t\leq t_2.\label{v2}\end{eqnarray}\end{theorem} Note that for a 100m, $\tau$ is between 1 and 2 $sec$ and $\alpha$
  is of order $10^{-2}$ $sec^{-1}$ so that $\alpha \tau<1$. We point out that $t_2$ is very close to
   $T$ and the values of $t_1$ and $t_2$ are determined by the
  condition on the distance and the energy equation as we will see below. The final value of the propulsive force is limited by how much aerobic energy can be provided by the runner since on the zero energy arc $f(t)=\sigma(0)/v(t)$.

\begin{proof}  We are going to use the properties of the switching function to determine the type of arcs.
 Indeed Theorem \ref{maint.t1} states that the trajectory starts with an arc of maximal force. We want to investigate
  what follows, whether the constraints are active and what the shape of the singular arc (no active constraint) is.

 We have that
Pontryagin's principle holds in qualified form,
i.e., each optimal trajectory
$(f,v,e)$ is associated with at least one multiplier
$(p,\mu)$ such that the relaxed control minimizes the
Hamiltonian.

The switching function is
\be
\Psi (t) = \frac{\partial H}{\partial f} (t)= p_v(t) - p_e(t) v(t) .
\ee We can differentiate this with respect to $t$ so that
 $\dot{\Psi} (t) =\dot{p}_v-\dot{p}_e v-p_e \dot{v}$, and we use
 (\ref{costeqaR}) to get,
\be\label{psipsi} \dot{\Psi}=1 + p_v /\tau+ p_e f+p_e \sigma'(e) v-p_e(f-v/\tau)=1+\Psi(t)/\tau+p_e v(\sigma'(e)+2/\tau).
\ee
We recall from Theorem \ref{maint.t1} that at $t=0$, the trajectory starts
on a maximal force arc and that,
 since $v=0$, $\Psi(0)=p_v(0)<0$. We want to investigate what can follow this arc of maximal force. Since
  we have the constraints $0\leq f \leq f_M$ and $e\geq 0$, there could be arcs of zero force, of zero energy,
  other arcs of maximal force or a singular arc.

{\em Step 1: Study of the singular arc.}
 On a singular arc, $\Psi (t) =0$.  We can differentiate this and find
 that
 $\dot{\Psi}=0$, so that, from (\ref{psipsi}), $ p_e v (\sigma'(e)+2/\tau)=-1$.
  In particular, on a singular arc, $\sigma'(e)+2/\tau \neq 0$.
 We differentiate again and obtain, after dividing by $p_e$, and using
 the equation for $\dot{e}$ and $\dot{v}$:
 \be\label{eqv}
\dot{v}\left (1-\frac{v^2\sigma ''(e)}{\sigma' (e)+2/\tau}\right )- v
\left(\sigma '(e)+\frac{\sigma ''(e)(\sigma (e)-v^2/\tau )}{\sigma' (e)+2/\tau}\right )=0.\ee
In a regime where $\sigma(e)=\alpha (e^0-e)$, the previous equation simplifies to
 (\ref{eqvsimp}).

 {\em Step 2: There is no zero force arc and hence,
$\Psi(t)\leq 0$, for all $t\in [0,T]$.} Let $(t_a,t_b)$ be a zero force arc, over which
necessarily $\Psi$ is nonnegative.
By what we have seen $t_a>0$,
and so $\Psi (t_a)=0$,  $\dot\Psi (t_{a+}) \geq 0$. On a zero force
arc, the equations yield that $p_e v=\lambda e^{(\alpha-1/\tau)t}$, hence is increasing since $\alpha \tau<1$ and $p_e(t)<0$. By \eqref{psipsi},
we have that
$$\dot \Psi(t_{b-}) - \dot \Psi(t_{a+})=\frac 1\tau \Psi(t_{b-})+(\frac 2 \tau -\alpha) ((p_ev )(t_{b-})
-(p_ev )(t_{a+})\geq 0$$
meaning that the zero force arc cannot end
before time $T$,
contradicting the final condition $e(T)=0$ recalled in Theorem \ref{maint.t1}.

{\em Step 3: The only maximal force arc is the one starting at $t=0$.} On a maximal force arc $(t_a,t_b)$
 with $t_a>0$,
 the speed increases  and $\dot{p}_e=\alpha p_e$, thus $p_e $ decreases and is negative. Therefore, $p_e v$ decreases.
 Equation \eqref{psipsi} implies
$\dot \Psi(t_{b-}) < \dot \Psi(t_{a+}) \leq 0$,
and since $\Psi\leq 0$ along the maximal force
arc, it follows that $\Psi(t_{b}) < 0$,
meaning that the maximal force arc ends at time $T$, which is in contradiction with $e(T)=0$ in Theorem \ref{maint.t1}.

{\em Step 4: end of the proof.}
The existence of a maximal force arc starting at time
0 is established in Theorem \ref{maint.t1}.
 Let $t_a\in (0,T)$ be its exit point.
Let
$t_b \in (0,T)$ be the first time at which the energy
vanishes.

If $t_a<t_b$, over $(t_a,t_b)$,
$\Psi$ is equal to zero and hence, $(t_a,t_b)$
is a singular arc on which (\ref{eqvsimp}) is satisfied. Finally let us show that
the energy is zero on $(t_b,T)$. Otherwise
there would exist $t_c,t_d$ with
$t_b \leq t_c < t_d \leq T$ such that
$e(t_c) = e(t_d) = 0$, and $e(t)>0$,
for all $t \in (t_c,t_d)$.
 Then $(t_c,t_d)$
is a singular arc, over which
$\dot e = \alpha(e^0-e) - fv$. We differentiate again and recall that $\dot{f}=-\alpha f$
 and $\dot{v}=-\alpha v$ on a singular arc, so that $$\ddot {e}=-\alpha\dot{e}+2\alpha fv.$$  Therefore, the function $e$
  cannot have a positive maximum
which gives a contradiction since the energy is positive.
The result follows.
 \end{proof}

  Equation (\ref{eqvsimp})  is the main result of the paper. It allows us to make the difference
 between short races and long races. Indeed, when the duration of exercise is less than 3mn, the maximal
 value of $\dot{VO2}$ is not reached, therefore $\sigma$ is a decreasing, almost linear function of energy and an increasing function
  of time, so that (\ref{eqvsimp}) provides that $v$ is a decreasing function of
  time once the runner can no longer run at maximal force.
  Therefors, the race starts at maximal force and $v$ is increasing. When on the singular arc,
   the velocity becomes
  decreasing and its evolution is governed by (\ref{eqvsimp}). This matches the model of \cite{sam15} of double exponential
  but provides a mathematical justification.

  We use (\ref{v1})-(\ref{v2}) to integrate the energy equation
  $$\frac{de}{dt}=\alpha (e^0-e)-fv$$ from 0 to $t_2$ recalling that $f=f_M$ for $t<t_1$ and that on
   the singular arc, we derive from (\ref{eqvsimp}) and (\ref{acc}) that $f=v(1/\tau -\alpha)$. We find
  \be\label{ent}\frac {e^0}{f_M^2\tau}e^{\alpha t_2}=\frac {e^{\alpha t_1}-1}\alpha-\frac
   {e^{(\alpha-\frac 1\tau) t_1}-1}{\alpha-\frac 1\tau}
  +(1-e^{-t_1/\tau})^2e^{2\alpha t_1}
  ( 1 -\alpha\tau)\frac {e^{-\alpha t_1}-e^{-\alpha t_2}}\alpha.\ee
  We also recall $D=\int_0^{t_2} v(t)\ dt$. In fact, there should be the last arc at zero energy, but since
  it is very short, we can make the approximation that the race is run in $t_2$. We find
  \be\label{dt1} \frac D {f_M\tau}=t_1-\tau (1-e^{-t_1/\tau})+(1-e^{-t_1/\tau})
  \frac {1-e^{-\alpha (t_2-t_1)}}\alpha.\ee This equation allows to solve for $e^{-\alpha t_2}$, and replace it in (\ref{ent}) to solve for $t_1$. The expression of $t_1$ is not analytic in general, but to get an idea on the dependance of the parameters, we can make assumptions. For instance, if we assume that $\alpha$ is small (of order $10^{-2}$ $sec^{-1}$), then we can expand the exponentials in $\alpha$ and find that (\ref{ent})-(\ref{dt1}) becomes
  \be\label{dt22} \frac D {f_M\tau}=t_1+ (1-e^{-t_1/\tau})(t_2-t_1-\tau),\ee
  \be\label{ent22}\frac D {f_M\tau}(1-e^{-t_1/\tau})- \frac {e^0}{f_M^2\tau}=(\tau -t_1-\tau e^{-t_1/\tau})e^{-t_1/\tau}.\ee Given the order of magnitude, the right hand side of (\ref{ent22}) can be neglected at leading order, which yields
  \be\label{eqt1}\frac D {f_M\tau}(1-e^{-t_1/\tau})- \frac {e^0}{f_M^2\tau}=0.\ee
   If $e^0=f_M D$, then $t_1$ is of order $t_2$, that is the race is run at maximal force and the $e^{-t_1/\tau}$  term can be in fact neglected,
   and one has to go further in the expansion in $\alpha$ of the exponentials. In general, $e^0$ is not so high, and $t_1$ is given by (\ref{eqt1}), that is
   \be\label{eqt12}t_1=-\tau \ln (1-\frac {e^0}{f_M D}).\ee Therefore, $t_1$ is an increasing function of $e^0$: the bigger the initial energy, the longer $f_M$ is maintained. Moreover, $t_1$ is a decreasing function of $f_M$: the bigger $f_M$, the shorter it can be maintained.


\hfill

 We have found that  the optimal race is to start at maximal force, and then decrease the propulsive force,
 because the runner does not have enough energy to maintain his maximal propulsive force for the whole duration
  of the race. The decrease in propulsive force leads to a singular arc and a decrease in velocity. It turns out that the
   decrease in the propulsive force predicted by our model is too strong compared to what the runner can produce. Therefore, in a real race, the runner keeps his maximal force
   for a shorter time and has a decrease which is weaker. This implies that in our model we have to take into
   account another constraint, which is the bound on the variation of the propulsive force. This implies also
   that in order to improve one's performance, improving the ability to vary the propulsive force
   is an important criterion.  This is why interval training is often used.

\subsection{Bounding variations of the force}
It seems desirable to avoid strong variations of the force which occur with the
previous model, and to introduce bounds on $\dot f$. The force
becomes then a state and the new control $\dot f$ is denoted by $g$.
So the state equation is
\be
\dot h=v;\quad \dot v = f - v/\tau;\quad \dot e= \sigma(e)  - fv;
\quad \dot f = g,
\ee
with constraints
\be
0 \leq f\leq f_M; \quad e\geq 0; \quad g_m \leq g \leq g_M.
\ee Note that $g_m$ is negative.
We minimize as before
$-\int_0^T v(t) \dd t$. The Hamiltonian is
\be
H = - v + p_v (f - v/\tau) + p_e(\sigma(e)-  fv) + p_f g.
\ee
The costate equation  are now
\be
\label{costeqb}
\left\{ \ba{lll}
-\dot p_v &= -1 - p_v /\tau - p_e f,
\\
-\dd p_e &= \sigma'(e) p_e \dd t - \dd\mu,
\\
- \dot p_f &= p_v - p_e v.
\ea\right.
\ee
The state constraint $e\geq 0$ is of second order,
and we may expect a
jump of the measure $\mu$ at time $T$.
The final condition for the costate are therefore
\be
p_v(T)=0; \quad p_e(T) = 0; 
\quad p_f(T)=0.
\ee
We may expect that the optimal
trajectory is such that $g$ is bang-bang (i.e., always on its bounds),
except if a state constraint is active (the state constraints now
include bound constraints on the force), as is confirmed in our
numerical experiments: for a 100m, the race starts with an arc of maximal force followed by an arc where
 $g=g_m$ and ends with an arc of zero energy.

The model with a bound on the derivative of the force matches the real race of world champions better
 than the previous system.
 This can be explained from the fact that the propulsive force cannot be varied too quickly from
 muscular and motor control reasons \cite{pes}. This is the model which is simulated to get Figure \ref{fig100M}.
  Analytically, if $\dot{f}=-\beta$, that is $f(t)=f_M-\beta (t-t_0)$,
  then the velocity equation can be solved explicitly:
  $$v(t)=v_0 e^{-(t-t_0)/\tau}+f_M \tau (1-e^{-(t-t_0)/\tau})+\beta \tau^2(1-e^{-(t-t_0)/\tau})-\beta \tau (t-t_0).$$
   So an approximation of the velocity close to $t_0$ is $f_M \tau-\beta(t-t_0)^2/2$.

   This formula can lead to an easier identification of the parameters: $f_M \tau$ is the peak velocity, $\beta $
    is related to
   the decrease of velocity and $\tau$ can be identified by matching the exponential at the beginning of the race.

\section{Longer races}
\begin{figure}
\centering
\begin{center}
\subfigure{\includegraphics [width=5.0cm]{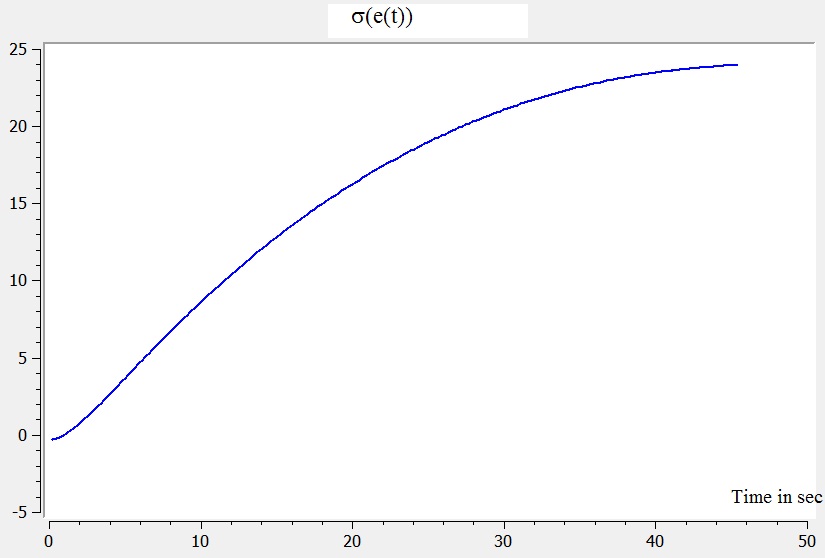}}
\quad
\subfigure{\includegraphics [width=5.0cm]{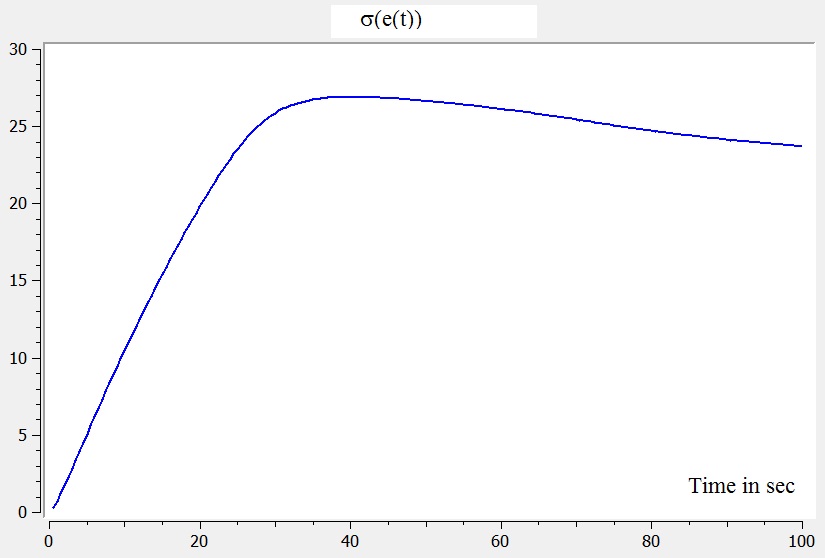}}
\end{center}\caption{$\dot{VO2}$ curve vs time, that is $\sigma(e(t))$ vs time for a 400m (left) and 800m (right).}
\label{figvo2}
\end{figure}
\begin{table}[!htb]
\caption{400m, French world level.
 Line 1 : distance in meter, line 2,  time splits for 50
meters in seconds. Total time 44.43 seconds.}\label{tab2}
\begin{tabular}{cccccccc}
\hline\\[-1.5ex]
 50m & 100m & 150m & 200m & 250m &  300m  &  350m & 400m \\[0.5ex]
\hline\\[-1.5ex]
6.10 & 4.94 & 5.0 &  5.17 & 5.37 & 5.52 & 5.86 & 6.45\\[0.5ex]
\hline
\end{tabular}
\end{table}
\begin{figure}
\centering
\begin{center}
\subfigure{\includegraphics [width=8.0cm]{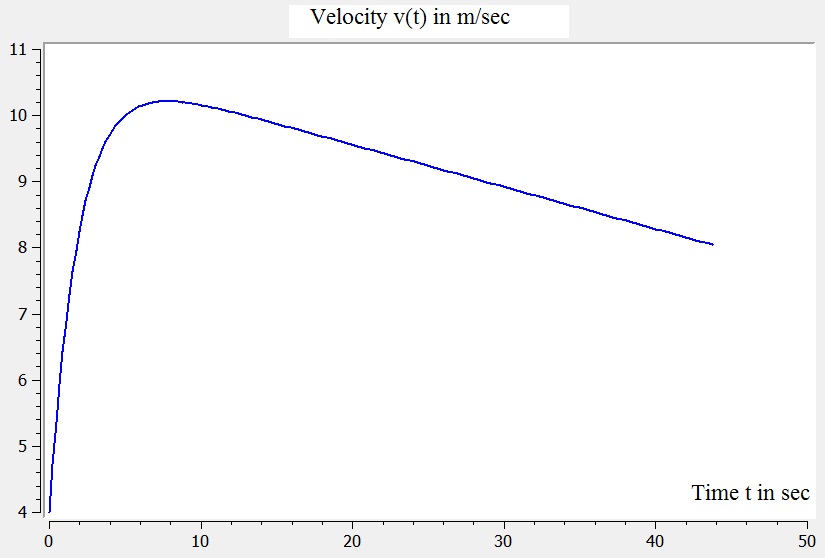}}
\quad
\subfigure{\includegraphics [width=10.0cm]{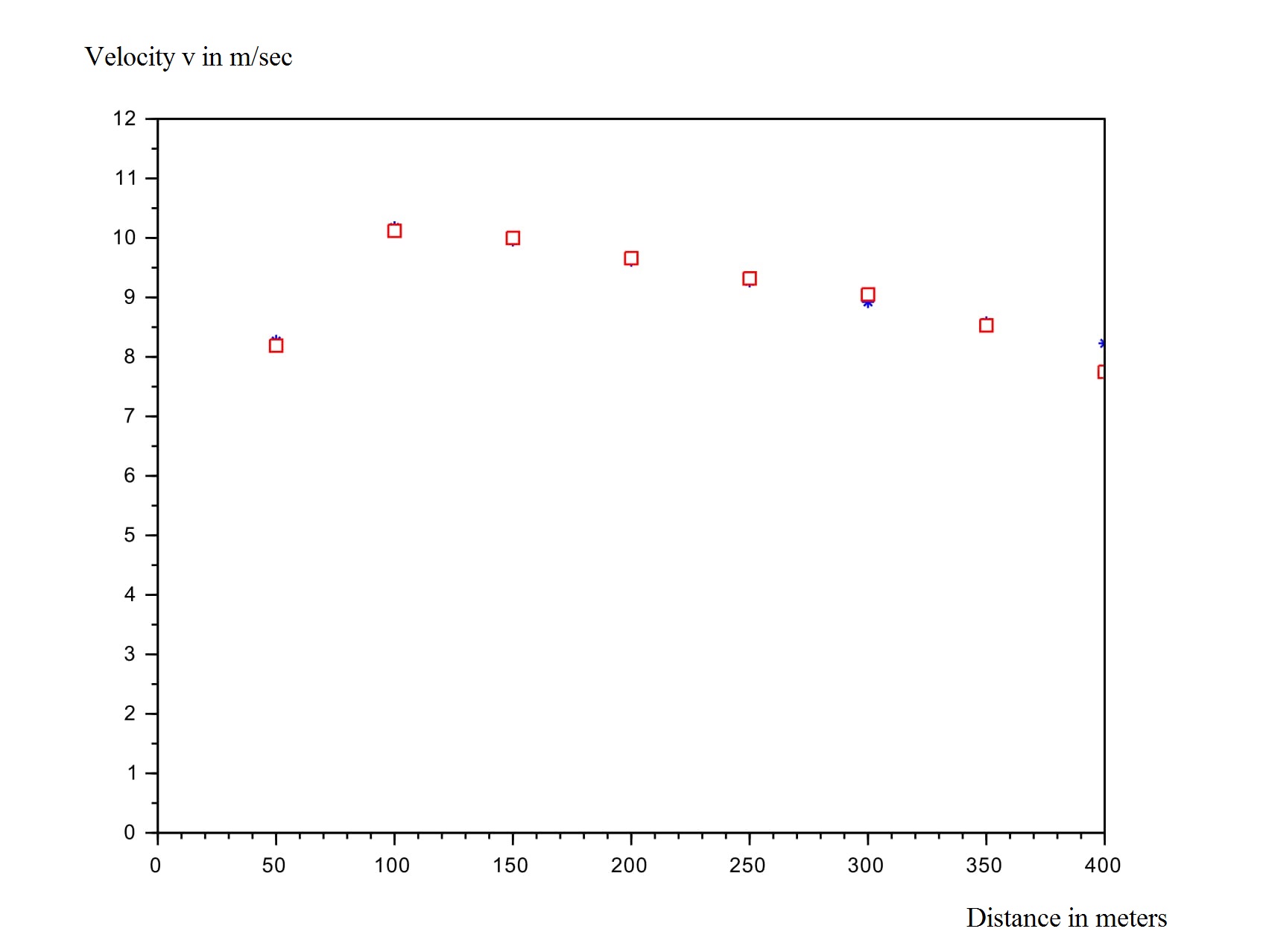}}
\end{center}\caption{Top: Plot of the velocity $v(t)$ vs time for a 400m. Bottom: every 50m,
 the numerical velocity is averaged (star)
 and compared to the mean value obtained from Table \ref{tab2}, as the ratio of 50 to the timesplit (square).
  The velocity is in
  $m.s^{-1}$ and is plotted vs the distance in m.}
\label{fig400}
\end{figure}

\begin{table}[!htb]
\caption{800m, Rudisha's timesplits for the 2012 Olympic games, World record and olympic record.
 Line 1 : distance in meter, line 2,  time splits for 100
meters in seconds. Total time 100.91 seconds.}\label{tab3}
\begin{tabular}{cccccccc}
\hline\\[-1.5ex]
 100m & 200m & 300m & 400m & 500m &  600m  &  700m & 800m \\[0.5ex]
\hline\\[-1.5ex]
12,3 & 11,2 & 12,55 & 13,23 & 12,74 & 12,28 & 13,02 & 13,59 \\[0.5ex]
\hline
\end{tabular}
\end{table}
\begin{figure}
\centering
\begin{center}
\subfigure{\includegraphics [width=8.0cm]{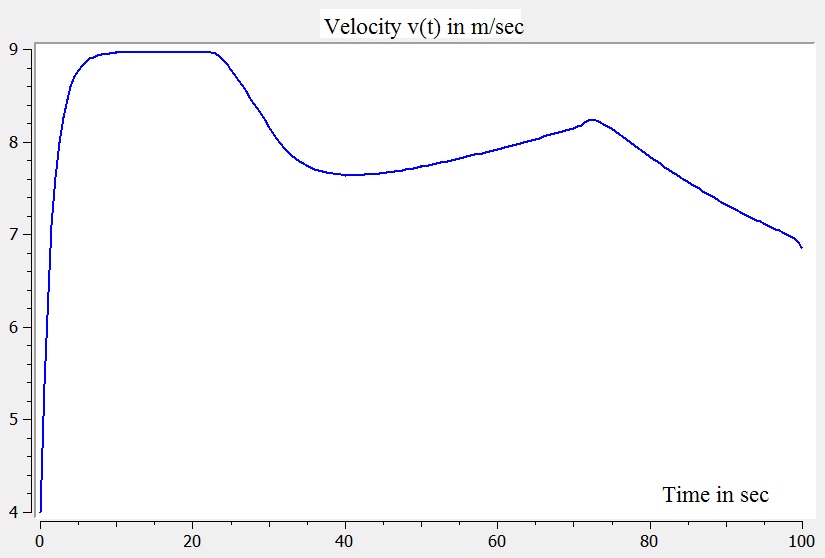}}
\quad
\subfigure{\includegraphics [width=10.0cm]{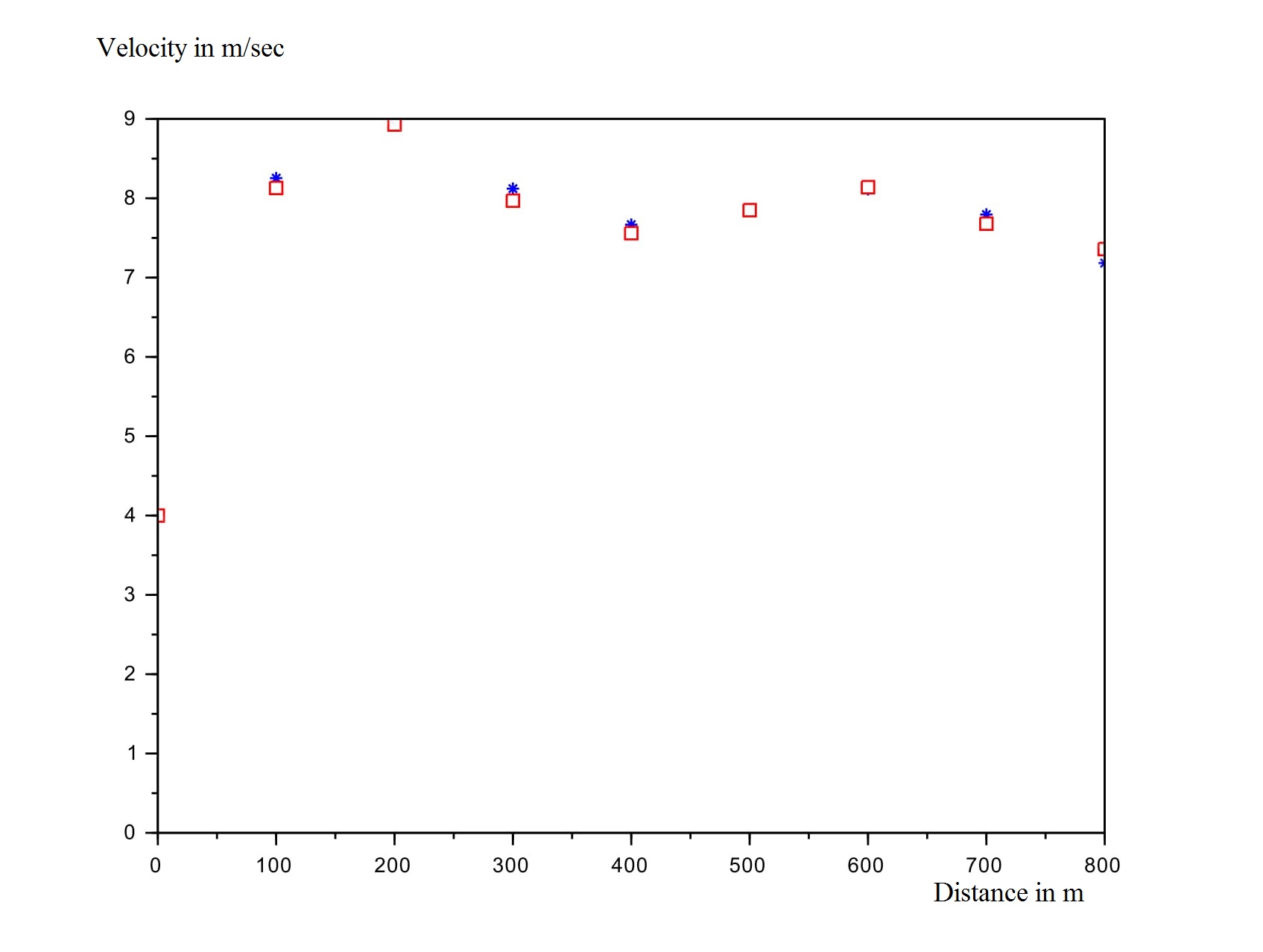}}
\end{center}\caption{Top: Plot of the velocity $v(t)$ vs time for an 800m. Bottom: every 100m,
 the numerical velocity is averaged (star)
 and compared to the mean value obtained from Table \ref{tab3}, as the ratio of 100 to the timesplit (square).
  The velocity is in
  $m.s^{-1}$ and is plotted vs the distance in m.}
\label{fig800}
\end{figure}
In order to describe mathematically a race, one needs  to know the $\dot{VO2}$ curve.
 We get information from \cite{hanoneffects} to construct the curve $\sigma(e)$. In Figure \ref{figvo2},
  we have plotted the $\dot{VO2}$ curve (that is $\sigma (e(t))$) vs time for a 400m and for an 800m.

  We have obtained data for Tables \ref{tab2} and \ref{tab3} from Christine Hanon \cite{han}
   as well as estimated data for $\bar \sigma$: for the 400m, it corresponds to French world level athletes,
   while for the 800m, it is the Olympic and World record of Rudisha (London 2012), where he was in front for the
    whole run, so that no strategy related to overtaking modified his way of running.

   In the 400m case, the $\dot{VO2}$ curve is
   increasing
  which leads to a velocity profile (Figure \ref{fig400} left) similar to a 100m, increasing and then
  decreasing
  though the decrease is stronger and the increase is only during the first quarter of the race, as can be confirmed
   in Table \ref{tab2}. The first part of the race
  is at maximal force, then the force is decreased and so is the velocity due to the fact that $\dot{VO2}$ is increasing. The description of the race of Theorem \ref{theomain} holds, as well as (\ref{eqt12}). { This is consistent with previous analyses in the literature leading to double exponentials for short races, but additionally provides a relation between the time where the runner starts slowing down and his total anaerobic energy.}

   In the  case of an 800m, the curve $\sigma(e(t))$ is increasing
  and then
   decreasing, which leads to a complicated race strategy: the beginning of the race is at maximal force and the
   velocity is increasing. {Then, the $\dot{VO2}$ is still increasing, and on the singular arc, the velocity is decreasing: indeed equation \eqref{eqvsimp} holds but $\alpha=\sigma' (e)>0$.
   At $t=40s$, the curve $\sigma(e(t))$ changes monotony and $\alpha$ changes sign and is negative again. Therefore, from  \eqref{eqvsimp}, the curve $v(t)$ is increasing again}
   until the runner reaches a situation where he can no longer increase his velocity because he does not have
   enough energy left and he moves onto a
   constrained arc where the derivative of the force is constant, before finishing on a zero energy arc.
   We have no mathematical proof that this is the best strategy in this order, but it is consistent with different
    possible arcs and the curve $\sigma(e(t))$. More precisely, the possibility of an arc of maximal force is allowed when $\sigma$ is an increasing function of energy, that is a decreasing function of time. In Theorem \ref{theomain}, { Step 3}, it was ruled out because $\sigma$ had the opposite monotony. We see from this part of the proof of Theorem \ref{theomain} that the decrease of velocity at the end of the race is related to the monotony of $\sigma$. Therefore,  if $\sigma$ changes monotony, as it is the case for longer races, then
 another maximal force arc is not excluded at the end of the race. This is why for longer races, at the end of the race,
  $\sigma$ is a decreasing function of time, and the velocity increases again: this is the final sprint.
  
  { Let us point out that this complex behaviour of the 800m race, had not been captured by other mathematical modeling: after the initial acceleration, Keller's model leads to a constant velocity and Ward Smith \cite{WS1} or Reardon \cite{reardon} to a decreasing velocity, while the reality of the race is more involved.}

When the duration of the race is longer than 3mn, the maximal value $\bar \sigma$ of
   $\dot{VO2}$
  is reached, $\sigma$ is a decreasing function of time at the end of the race, or an increasing linear function of energy,
  and therefore, (\ref{eqv}) implies that
  $v$ is increasing again.

 Varying the parameters, we can notice that a high $\dot{VO2}$ (high $\bar \sigma$)
  improves performance as well as a high ability to vary the propulsive force. Hence
 interval training exercises are favored to improve this force variation.

\begin{figure}
\centering
\begin{center}
\subfigure{\includegraphics [width=6.0cm]{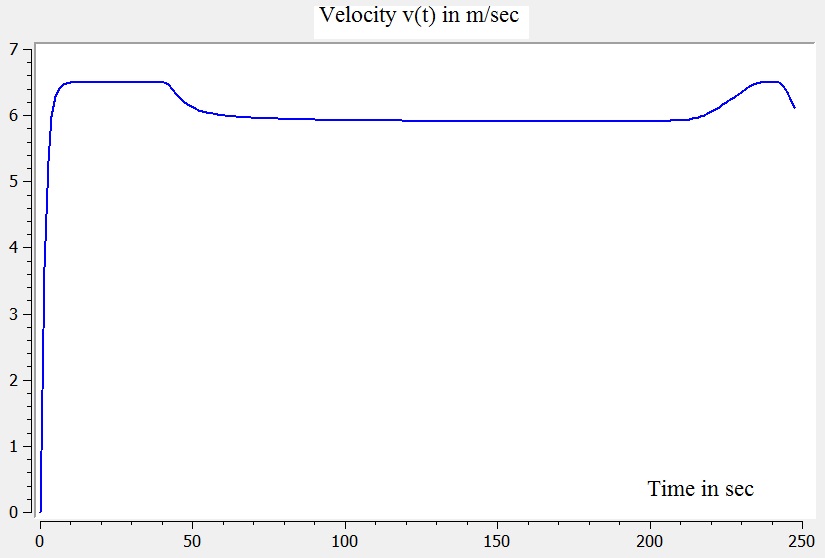}}
\quad
\subfigure{\includegraphics [width=6.0cm]{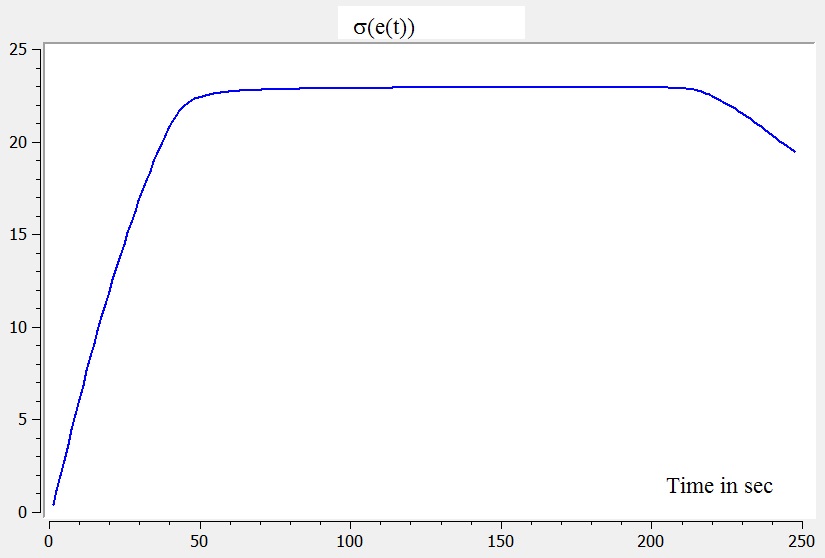}}
\end{center}\caption{Velocity curve vs time for a 1500m (left) and $\dot{VO2}$ curve, that is $\sigma(e(t))$
  vs time (right).}
\label{fig1500}
\end{figure}
The 1500m has been studied in \cite{AB} and \cite{aftaction}, in particular the effects of the various parameters.
 Here, we use the data of \cite{hanonpacing} to produce the velocity curve and $\dot{VO2}$ curve in Figure
  \ref{fig1500}. Again, we can derive information from (\ref{eqv}): when in the part where $\dot{VO2}$ is linearly increasing,
   on the singular arc, we have (\ref{eqvsimp}) with $\alpha >0$ and the
   velocity is decreasing; on the part where $\sigma$ is constant, the velocity is constant, and when $\sigma$ is
    linearly decreasing, we have (\ref{eqvsimp}) with $\alpha <0$ and the velocity is increasing so that there is a final sprint. The very last points where the velocity decreases
    again is due to the zero energy part at the end of the optimization.
    
    For longer races, we do not have detailed timesplits but we believe that the model remains valid, and could provide interesting indications in the case of hilly and windy races.


\section{Conclusion}  Given a distance $D$ to run, our model relies on the knowledge of the $\dot{VO2}$ curve, that is the data of $\bar \sigma$, $\sigma_f$, $\lambda$ and $e_{crit}$ and on 3 parameters: the maximal propulsive force $f_M$, the initial anaerobic energy $e^0$ and the friction coefficient $\tau$. On the basis of these parameters, the energy conservation and Newton's second law, we can determine the optimal way of running, that is the instantaneous velocity, propulsive force and anaerobic energy. The same model holds for all distances { but provides very different shapes of velocities according to distances.}

The structure of the 100m or 400m race is that $v$ increases exponentially quickly when $f$ is at its maximal value
 $f_M$.
 Then $v$ decreases exponentially slowly until the energy reaches the zero value for a few points of computation. { Though the double exponential profile structure for the velocity was already known in the literature, a precise dependance on the different parameters is clearly identified.}
  One sees that in order to reach a very high velocity quickly, one needs a very strong propulsive force ($f_M$)
   and small internal friction dues to joints or running economy (high $\tau$). But on the other hand,
    the limiting value becomes the available anaerobic energy because reaching a high velocity quickly requires
    a lot of energy, and if little is left, the runner cannot maintain his velocity, which falls down exponentially.
    Nevertheless, the best strategy is to put as high force and acceleration at the beginning to reach strong velocity,
    even if this strong velocity cannot be maintained, rather than accelerating more slowly and all along the race.
    
    If the race is long enough so that $\sigma$ reaches its maximal value and decreases on the last third of the race, then the runner can speed up again. { We provide a precise structure of the $800$m race, for which the velocity curve has 4 pieces: increasing, decreasing, increasing and decreasing again.} For a $1500$m or more, there is a long part of the race at almost constant velocity, which gets close to Keller's model.
      Numerically, our
  model matches 100m, $400$m, $800$m and $1500$m races for olympic or world championships. 

    The interest of the mathematical model is that it allows to play on the parameters variations. For instance,
    one can identify how the variation of one parameter or another has an influence on the deceleration on the second part
    of the race or how to better improve the final time since our model is strong enough to match
    all type of distances and human optimization.

\bigskip

\paragraph{\bf Acknowledgments}
The author would like to thank her colleague  Martin Andler, who is both a talented mathematician
and a former 800m runner, and whose remarks and advice all along this work were crucial. Martin Andler introduced
the author to a physiologist and former French champion Christine Hanon. Christine Hanon provided all real race data
 for 400, 800 and 1500m \cite{han}. Though she participated strongly to this
work, she did not want to be involved in the authorship. Her involvement into making physiologists
and sport scientists believe in mathematics is strongly acknowledged here, as well as her terrific ability to understand
mathematics. The author is also very grateful to
 Fr\'ed\'eric Bonnans  and Pierre Martinon on Bocop and to people from the sprint project at
 Insep, namely Antoine Couturier, Gael Guilhem and Giuseppe Rabita who provided the timesplits of Table \ref{tab1}.

\bibliographystyle{plain}

\bibliography{bibliosport,bibliosport2}

\end{document}

%% file: macro.tex
\def\dd{{\rm d}}

\def\weight(#1,#2){c_{#1,#2}}

\if {

}{{\caption}}
} \fi

\def\tt{\tilde{t}}

\def\1B{{\bf  1}}

\def\supp{\mathop{\rm supp}}

\def\Min{\mathop{\rm Min}}

\def\1B{{\bf  1}}

\newcommand\be{\begin{equation}}
\newcommand\ee{\end{equation}}
\newcommand\ba{\begin{array}}
\newcommand\ea{\end{array}}
\newcommand{\bea}{\begin{eqnarray*}}
\newcommand{\eea}{\end{eqnarray*}}
\newcommand{\bean}{\begin{eqnarray*}}
\newcommand{\eean}{\end{eqnarray*}}

